\documentclass[twocolumn,showpacs,preprintnumbers,amsmath,amssymb]{revtex4}

\usepackage{graphicx}
\usepackage{dcolumn}
\usepackage{bm}

\begin{document}

\preprint{APS/123-QED}

\title{Superconductivity Induced by the Proximity Effect
of Singlet \\Resonating Valence Bond 
Order in High-$T_c$ Superconductors}

\author{Masahiko Hayashi}
 \email{hayashi@cmt.is.tohoku.ac.jp}
\author{Hiromichi Ebisawa}%
\affiliation{%
Graduate School of 
Information Sciences, Tohoku University, \\
Aramaki Aoba-ku, Sendai 980-8579, Japan
}%

\date{\today}

\begin{abstract}
Effects of inhomogeneous doping on the high-$T_c$ cuprate 
superconductors are studied within the 
framework of the $t$-$J$ model. 
Especially, the boundary between two non-superconducting regions 
with doping rates much higher and lower than the 
optimal one is examined. 
It is found that, although there 
is no superconductivity in the bulk, 
a superconducting region appears at the boundary 
because singlet resonating valence bond order 
and holon condensation occur simultaneously in this region. 
The critical temperature of the induced superconductivity 
can be higher than that of the optimally doped sample and 
the critical current density can be the same order 
as that in the superconducting state 
below the critical temperature. 
We also point out an experimental possibility to 
observe this phenomenon. 
\end{abstract}

\maketitle

It has been recognized 
that there are two important ingredients 
for the superconductivity in 
high-$T_c$ cuprates, which are the 
spin correlation, leading to the spin gap, 
and the doped charge carriers with a sufficiently high density. 
In the slave-boson mean-field 
theory of the $t$-$J$ model, the former is realized as 
the singlet resonating valence bond (s-RVB) 
order of spinons and the latter the bose condensation of holons 
\cite{Anderson,Baskaran-Zou-Anderson,%
Anderson-Zou,Isawa-Maekawa-Ebisawa,%
Anderson-Baskaran-Zou-Hsu,%
Suzumura-Hasegawa-Fukuyama,Kotlier-Liu}. 
In this framework, it is assumed that 
the spinons and the holons are 
well-defined low lying excitations, 
and the spin-charge separation is a good picture. 
Although there still remains some controversy on this point, 
it has been clarified that some 
physical properties of the high-$T_c$ 
cuprates can be well described by the $t$-$J$ model, 
for example, the transport properties 
\cite{Nagaosa-Lee} and the magnetic properties 
\cite{Tanamoto-Kuboki-Fukuyama,%
Tanamoto-Kohno-Fukuyama}. 
In this Letter, we take the $t$-$J$ model to be our starting 
point. 

We consider the situation 
where the doping rate is spatially varying. 
We especially concentrate on the case where 
two non-superconducting regions, 
one is too over-doped and the other is too 
under-doped, 
are in contact with each other within the 
same CuO$_{2}$ plane. 
We expect that
the s-RVB order and the holon condensate 
can exist simultaneously at the boundary, 
thus giving rise to 
superconductivity. 
Since the s-RVB order and the holon condensation 
can occur at temperatures higher than 
the critical temperature of the optimally doped sample, 
$T_{\rm opt}$, 
this \lq\lq boundary superconductivity\rq\rq 
may also occur above $T_{\rm opt}$. 

Our idea is similar with the \lq\lq spin-gap 
proximity effect\rq\rq previously introduced by Emery, Kivelson 
and Zacher \cite{Emery-Kivelson-Zacher} based on a 
somewhat different model for high $T_{c}$ superconductors, 
discussing the stripe order and its significance for the 
superconductivity. 
Recently, Kivelson has proposed a possibility 
to raise critical temperature using spin-gap 
proximity effect \cite{Kivelson}. 
He proposes making a stack of spin-gap material and 
hole-rich material in a layered structure. 
In contrast to this, we consider the situation 
where the spin-gap and hole-rich regions 
exist within a CuO$_{2}$ plane. 
This kind of situation has not been studied 
within the framework of the $t$-$J$ model and 
we consider this is of interest 
from both theoretical and experimental 
points of view. 

First we introduce the $t$-$J$ model which 
describes the correlated electrons in each CuO$_{2}$ plane. 
The Hamiltonian is given by 
\begin{eqnarray}
    {\cal H} &=& -t \sum_{\langle i,j\rangle \sigma}
    (f_{i \sigma}^{\dagger}f_{j \sigma} b_{j}^{\dagger}
    b_{i} + {\rm h.c.})+
    J\sum_{\langle i,j\rangle}{\vec S}_{i}\cdot{\vec S}_{j}
    \nonumber\\
    && - \sum_{i}\lambda_{i}(
    \sum_{\sigma}f_{i \sigma}^{\dagger}f_{i \sigma}+
    b_{i}^{\dagger}b_{i}-1),
    \label{Hamiltonian}
\end{eqnarray}
where $i$ and $j$ denote lattice points
in the CuO$_{2}$ plane, 
$\langle i,j\rangle$ means the 
nearest neighbors, 
$f_{i \sigma}$ ($\sigma=\uparrow,\downarrow$) 
and $b_{i}$ are the annihilation operators of 
the spinon and the holon, respectively, 
${\vec S}_{i}$ denotes 
$\frac{1}{2}\sum_{\alpha\beta}
f_{i \alpha}^{\dagger}{\vec\sigma}_{\alpha\beta}f_{i \beta}$ 
with ${\vec \sigma}_{\alpha\beta}$ being Pauli matrices, and 
$\lambda_{i}$ is the Lagrange multiplier for the 
constraint, $\sum_{\sigma}f_{i \sigma}^{\dagger}f_{i \sigma}+
b_{i}^{\dagger}b_{i}=1$. 
In this Letter the lattice spacing is taken to be unity. 
In addition to Eq. (\ref{Hamiltonian}), we later introduce the 
coupling to the electromagnetic field. 

The Eq. (\ref{Hamiltonian}) is treated by the so-called 
slave-boson mean-field theory. 
The local constraint represented by 
$\lambda_{i}$ is replaced by the global one, 
which is described by 
the constant chemical potential 
for spinons $\mu_{\rm F}$ and that for holons $\mu_{\rm B}$. 
We define the statistical averages, 
$\xi \equiv \langle \sum_{\sigma}f_{i \sigma}^{\dagger}
f_{j \sigma}\rangle$ and 
$\eta \equiv \langle b_{i}^{\dagger}b_{j}\rangle$, 
which are assumed to be independent of 
$i$ and $j$ \cite{average}. 
The order parameter for the s-RVB order is denoted as 
$\Delta_{ij}=\frac{3J}{8}\langle 
f_{i\uparrow}f_{j\downarrow} - f_{i\downarrow}f_{j\uparrow}
\rangle$, which is equal to $\Delta_{x}({\vec r}_{i})$ 
if $j=i+{\hat x}$ and 
$\Delta_{y}({\vec r}_{i})$ if $j=i+{\hat y}$ where 
${\hat x}$ and ${\hat y}$ indicate 
the shift by one lattice spacing in positive 
$x$- and $y$-direction, respectively. 
Although $\xi$, $\eta$ and $\Delta_{ij}$ should be 
determined self-consistently in principle, 
we skip this process for simplicity by just 
assuming $\xi$ and $\eta$ to be constant. 
Including also the scalar and vector potential, 
$\varphi$ and ${\vec A}$, and the fictitious 
gauge fields, $a_{0}$ and ${\vec a}$, which represent 
the fluctuation around the mean-field solution 
\cite{Nagaosa-Lee}, 
we obtain the following Lagrangian, 
\begin{eqnarray}
    {\cal L} &=& \sum_{i\sigma}f_{i\sigma}^{\dagger}
    \left\{\hbar\partial_{\tau}-\mu_{\rm F}+i a_{0}({\vec r}_{i})
    \right\}f_{i\sigma} 
    \nonumber\\
    && + \sum_{i}b_{i}^{\dagger}
    \left\{\hbar\partial_{\tau}-\mu_{\rm B}+i 
    (e \varphi({\vec r}_{i})+a_{0}({\vec r}_{i}))
    \right\}b_{i}
    \nonumber\\
    && + \sum_{i} i \varphi({\vec r}_{i}) \rho({\vec r}_{i})
    \nonumber\\
    && - \left(t \eta + \frac{3}{8} J \xi\right)
    \sum_{\langle ij\rangle \sigma}\left\{
    {\rm e}^{i \theta^{\rm F}_{ji}}f_{i\sigma}^{\dagger}
    f_{j\sigma}+{\rm h.c.}
    \right\}
    \nonumber\\
    && - t \xi \sum_{\langle ij\rangle}
    \left\{
    {\rm e}^{i \theta^{\rm B}_{ji}}b_{i}^{\dagger}
    b_{j}+{\rm h.c.}
    \right\}
    \nonumber\\
    && - \sum_{i}\biggl[
    \Delta_{x}^{*}({\vec r}_{i}) \chi_{i i+{\hat x}} + 
    \Delta_{y}^{*}({\vec r}_{i}) \chi_{i i+{\hat y}} +
    {\rm h.c.}
    \nonumber\\
    && 
    \phantom{aaaaaa}
    - \frac{8}{3J} (|\Delta_{x}({\vec r}_{i})|^2 +
    |\Delta_{y}({\vec r}_{i})|^2)
    \biggr], 
    \label{Lagrangean}
\end{eqnarray}
where $\chi_{ij} = f_{i\uparrow}f_{j\downarrow}-
f_{i\downarrow}f_{j\uparrow}$ and 
\begin{equation}
    \theta^{\rm B}_{ji} = \frac{1}{\hbar}
    \int_{{\vec r}_{j}}^{{\vec r}_{i}}
    {\vec a}\cdot{\rm d}{\vec l},\,\,\,
    \theta^{\rm F}_{ji} = \frac{1}{\hbar}
    \int_{{\vec r}_{j}}^{{\vec r}_{i}}
    \left({\vec a}+\frac{e}{c}{\vec A}
    \right)\cdot{\rm d}{\vec l}. 
\end{equation}
The chemical potentials $\mu_{\rm F}$ and 
$\mu_{\rm B}$ are determined from 
\begin{equation}
\langle f_{i\sigma}^{\dagger}
f_{i\sigma}\rangle = 1 - {\bar \delta},\,\,
\langle b_{i}^{\dagger}
b_{i}\rangle = {\bar \delta}, 
\end{equation}
where ${\bar \delta}$ is the average
holon density in the whole system. 
The background charge density, originating from doping, 
is denoted by $\rho({\vec r}_{i})$, 
whose spatial average equals $-e{\bar \delta}$. 
In this Letter $a_{0}$ and ${\vec a}$ 
are treated perturbatively 
on the same line with Refs. \cite{Nagaosa-Lee} 
and \cite{Nagaosa-Lee2}. 

First we study how the distribution 
of the holons, namely $\delta({\vec r}_{i})$, 
is determined under an inhomogeneous doping, 
which is described by $\rho({\vec r}_{i})$. 
For this purpose, we neglect the terms including 
$\Delta_{ij}$ and ${\vec a}$ and ${\vec A}$. 
Then the spinon and the holon degrees of freedom 
can be integrated out, giving rise to terms, 
$\pi_{\rm F} a_{0}^2 + \pi_{\rm B} (a_{0}+e \varphi)^2$, 
where $\pi_{\rm F}$ and $\pi_{\rm B}$, respectively, 
are the polarization 
functions of spinons and holons. 
Now the magnitudes of 
$\pi_{\rm F}$ and $\pi_{\rm B}$ 
determine which of the spinon and the holon couples to 
$\varphi$. 
In this Letter we assume 
$\pi_{\rm F} \gg \pi_{\rm B}$, 
which may be valid in lower doping region. 
Then $a_{0}$ is almost fixed to zero and 
$\varphi$ couples to the holons. 
Now $\delta$ and $\varphi$ are determined 
from $\rho$ in a self-consistent way. 

\begin{figure}
\includegraphics[width=5cm,clip]{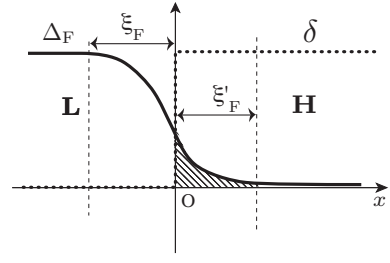}
\caption{Spatial variation of 
     the holon density $\delta$ (dotted line) and the s-RVB order parameter 
    $\Delta_{d}$ (bold line).
    {\bf H} and {\bf L} stand for high and 
    low doping region, respectively.
    In the shaded region the superconductivity is 
    expected to appear. }
    \label{s-RVB_holon-density}
\end{figure}

We especially consider 
the following situation. 
The CuO$_{2}$ planes, described by the above 
Lagrangian, are laid 
parallel to $x$-$y$ plane and 
stacked in $z$-direction. 
The region $x < 0$ ($x > 0$) is 
filled with a s-RVB ordered (hole-rich) material with a  
low (high) enough doping rate. 
Both of the regions are non-superconducting. 
The system is uniform in $y$- and $z$-direction. 
We assume that the doping rate, 
which is described by $\rho({\vec r}_{i})$, 
changes at $x=0$ abruptly. 
In order to preserve the charge neutrality 
in the bulk, namely at 
$x \rightarrow \pm \infty$, 
the statistical average 
$\langle\varphi({\vec r})\rangle$ must 
take constant imaginary values, 
which differ between $x \rightarrow \infty$ and 
$x \rightarrow -\infty$. 
(Note that the chemical potential $\mu_{\rm B}$ 
is a constant in the whole system.) 
Near the boundary 
$\langle\varphi({\vec r})\rangle$ 
changes smoothly connecting these two bulk limits. 
As in the case of semiconductor 
junctions \cite{Sze}, 
the smooth connection is enabled by the polarization charge 
appearing at the boundary. 
The depth of the charged region is given by 
the screening length of the holons. 
This length is extremely long (short) 
for $x < 0$ ($x > 0$) and 
there still remains a sharp drop of the holon density 
at the boundary. 
Therefore, in this Letter, we approximate 
$\delta$ by a step function 
(see Fig. \ref{s-RVB_holon-density}). 

Next we study the s-RVB order by 
introducing the Ginzburg-Landau (GL) 
free energy of the order parameter 
$\Delta_{d} = \frac{1}{2}(\Delta_{x}-\Delta_{y})$. 
Here we assume that the s-RVB order has $d$-wave symmetry. 
The GL free energy is obtained by the perturbative expansion 
of the free nergy with respect to $\Delta_{d}$ \cite{Feder-Kallin}. 
We introduce the half band width $D = 4 ( t\eta + 3 J \xi/8)$ and 
the strength of attractive interaction $V = 3J/8$. 
The dispersion of spinons,  
$(D/2)(\cos k_{x} + \cos k_{y})$, is approximated 
by a parabolic one, 
$D (-1 +|{\vec k}|^2/4)$, and then 
the density of states at Fermi energy becomes 
$N(0) = 1/(\pi D)$. 
At a temperature $T \ll D$, the solution for  
$\sum_{\sigma}
\langle f_{i \sigma}^{\dagger}f_{i \sigma}\rangle 
=1 - \delta$
is approximately given by 
$\mu_{\rm F} = -\delta /N(0)$ and 
the critical temperature $T_{d}$ becomes 
$T_{d} \simeq T^{*}\exp[-1/\{2 V N(0) (1-\delta)^2\}]$ 
with $T^{*} = 2 D {\rm e}^{\gamma}/\pi$ where $\gamma$ 
is the Euler's constant. 
This expression is valid in the 
\lq\lq weak coupling regime\rq\rq, 
$V N(0) \ll 1$. 
The GL free energy density of $\Delta_{d}$ is given by 
\begin{eqnarray}
    F_{\rm s-RVB}=F_{0}+\alpha_{d} |\Delta_{d}|^2
    + \beta |\Delta_{d}|^4 
    + \gamma_{d}|{\vec \Pi}\Delta_{d}|^2,
    \label{GL}
\end{eqnarray} 
where ${\vec \Pi} = -i \nabla +(2/\hbar) {\vec a} +
(2\pi/\phi_{0}){\vec A}$
with $\phi_{0} = hc/(2e)$. 
The GL coefficients are given by 
\begin{eqnarray}
    \alpha_{d} &=& 2 N(0) (1-\delta)^2 \ln\frac{T}{T_{d}},
    \\
    \beta &=& \frac{21 \zeta(3) N(0)}{2 \pi^2 T^2}
    (1-\delta)^4\equiv 
    \frac{c_{1}}{T^2 D},
    \\
    \gamma_{d} &=& 
    \frac{7 \zeta(3) N(0) D^2}{16 \pi^2 T^2}\equiv 
    \frac{c_{2} D}{T^2},
\end{eqnarray}
where $\zeta(x)$ is the zeta function and $c_{1}$ and 
$c_{2}$ are constants. 
The doping dependence of the GL coefficients comes from 
$T_{d}$, $\alpha_{d}$ and $\beta$. 
Since the largest dependence appears from $T_{d}$, 
we neglect the $(1-\delta)^{n}$-factors in 
$\alpha_{d}$ and $\beta$ in the following. 
Then the doping rate affects $T_{d}$ only. 

Here we consider the situation 
where the transition temperatures 
in the regions $x<0$ and $x>0$, 
denoted by $T_{d}$ and $T'_{d}$, 
respectively, 
satisfies the condition, 
$T_{d} > T \gg T'_{d}$. 
It is also assumed that holons are condensed in 
the region $x>0$. 
Note that the quantity with (without) 
dash ($'$) is defined in $x > 0$ ($x < 0$). 
We first determine the spatial variation 
of $\Delta_{d}$, disregarding 
${\vec a}$ and ${\vec A}$. 
Here $\Delta_{d}$ is assumed to be real. 
The expectation value of $\Delta_{d}$ 
is $\sqrt{|\alpha_{d}|/(2 \beta)} \equiv 
{\bar \Delta_{d}}$ at $x \rightarrow -\infty$
and $0$ at $x \rightarrow \infty$. 
In $x<0$ we introduce a new variable 
$\delta \Delta_{d} = \Delta_{d} - {\bar \Delta_{d}}$. 
The spatial variation of $\Delta_{d}$ is then 
governed by the following equations, 
\begin{eqnarray}
    && 2 |\alpha_{d}| \delta\Delta_{d}(x) - 
    \gamma_{d}\partial_{x}^{2} \delta\Delta_{d}(x)
    =0
    \phantom{aaaaa}
    (x<0),
    \nonumber\\
    && \alpha'_{d} \Delta_{d}(x) - \gamma_{d} 
    \partial_{x}^{2} \Delta_{d}(x)
    =0 \phantom{aaaaa} (x>0).
\end{eqnarray}
We can easily see that $\Delta_{d}$ behaves as 
${\bar \Delta_{d}} + g_{1} \exp(x/\xi_{\rm F})$ for $x<0$ 
and $g_{2} \exp(-x/\xi'_{\rm F})$ for $x>0$, 
where $\xi_{\rm F}=\sqrt{\gamma_{d}/2|\alpha_{d}|}$, 
$\xi'_{\rm F}=\sqrt{\gamma_{d}/\alpha'_{d}}$, 
and $g_{1}$ and $g_{2}$ are constants. 
The solution in the whole region 
is obtained by connecting these two solutions at $x=0$ 
with boundary condition, 
\begin{equation}
    \partial_{x}\Delta_{d}(-0) = \partial_{x}\Delta_{d}(+0),\,\,
    \Delta_{d}(-0) = \Delta_{d}(+0), 
\end{equation}
(see Fig. \ref{s-RVB_holon-density}). 
This condition is in contrast with that for 
the proximity effect in ordinary 
superconductor-normal metal junction, where the change of 
the BCS interaction causes a discontinuity of the gap function
\cite{deGennes}. 
(Note that the quantity which is continuous at the 
boundary is not the gap function but the anomalous amplitude.) 
The constants $g_{1}$ and $g_{2}$ are obtained as  
\begin{equation}
    g_{1}=\frac{\xi_{\rm F}}{\xi_{\rm F}+\xi'_{\rm F}}
    {\bar \Delta}_{d}, 
    \,\,\,
    g_{2}=\frac{\xi'_{\rm F}}{\xi_{\rm F}+\xi'_{\rm F}}
    {\bar \Delta}_{d}.
\end{equation}
Note that non-zero $\Delta_{d}$ in $x>0$ is 
due to the proximity effect of s-RVB order and 
$\xi'_{\rm F}$ gives the proximity length. 
The temperature dependence of the coherence 
lengths at $T'_{d} \ll T \alt T_{d}$ are as follows:  
\begin{eqnarray*}
    \xi_{\rm F} 
    \simeq \sqrt{\frac{c_{2}}{2}}\frac{D}{T_{d}}
    \frac{1}{\sqrt{1-T/T_{d}}},\,\,\,
    \xi'_{\rm F} = 
    \frac{\sqrt{c_{2}}}{\sqrt{\ln T/T'_{d}}}
    \frac{D}{T}.
\end{eqnarray*}
It is interesting to see that $\xi'_{\rm F}$ is, 
except for the logarithmic factor, 
analogous to the ordinary proximity length in the clean limit 
$\propto \hbar v_{\rm F}/T$ where $v_{\rm F}$ 
is the Fermi velocity. 
If $T$ is close to $T_{d}$, 
$\xi_{\rm F}$ becomes much larger than $\xi'_{\rm F}$ 
and the amplitude of $\Delta_{d}$ in the 
proximity region, {\it i.e.}, $g_{2}$, becomes much smaller 
than ${\bar \Delta}_{d}$. 
However if $T$ is not too close to $T_{d}$, 
$\xi_{\rm F}$ and $\xi'_{\rm F}$ can be the same order and 
$g_{2}$ becomes comparable to ${\bar \Delta}_{d}$. 
Noting that the holons are 
already bose condensed in the proximity region, 
this result means the induced superconductivity in the 
boundary area. 

The strength of the induced superconductivity 
becomes more apparent 
by estimating the supercurrent which can flow parallel 
to the boundary. 
In this region the gauge field ${\vec a}$ is massive 
because of the holon condensation, and 
${\vec a}$ in Eq. (\ref{GL}) can be set to zero. 
Then the response to the actual vector potential ${\vec A}({\vec r})$
is dominated by the spinons. 
Here we take the vector potential to be constant 
and parallel to $y$-direction. 
The total current in $y$-direction is 
given by 
\begin{eqnarray}
    J &=& 2 c \gamma_{d} A_{y} \int_{0}^{\infty}
    \Delta_{d}^2(x)\, {\rm d}x
    \nonumber\\
    &=& c \gamma_{d} A_{y}
    \frac{\{{{\tilde\xi}'_{\rm F}} (A_{y})\}^3}
    {\{{\tilde\xi}'_{\rm F}(A_{y})+\xi_{d}\}^2}
    {\bar \Delta}_{d}^2,
    \label{current}
\end{eqnarray}
where the $A_{y}$-dependent 
proximity length ${\tilde\xi}'_{\rm F}(A_{y})
=\xi'_{\rm F}/\sqrt{1+p' A_{y}^2}$ 
decreases with increasing $A_{y}$, 
with $p'$ being $(2 \pi \xi'_{\rm F}/\phi_{0})^2$. 
The maximum of $J$ as a function of $A_{y}$ is given 
approximately by 
\begin{equation}
    J_{\rm max}=2c \left(\frac{2\pi}{\phi_{0}}\right)^2
    \gamma_{d}\frac{1}{3\sqrt{3p'}}
    \frac{{\xi'_{\rm F}}^3}{{\xi_{\rm F}}^2}
    {\bar \Delta}_{d}^2. 
\end{equation}
For clarity, here we introduce the 
critical current density $j_{c}$ 
in the bulk superconducting phase. 
This quantity (in overdoped region) 
is calculated from Eq. (\ref{GL}) 
by assuming that holons are condensed and 
${\vec a}$ is massive in the whole region. 
We obtain 
\begin{equation}
    j_{c}=2c \left(\frac{2\pi}{\phi_{0}}\right)^2
    \gamma_{d}\frac{1}{3\sqrt{3p}}
    {\bar \Delta}_{d}^2,
\end{equation}
where $p = 2 (2 \pi \xi_{\rm F}/\phi_{0})^2$. 
It is clear that $J_{\rm max}= 
\sqrt{2} j_{c} {\xi'_{\rm F}}^{2}/\xi_{\rm F}$. 
If $\xi'_{\rm F} \simeq \xi_{\rm F}$, 
the critical current density 
in the proximity region, defined by $J_{\rm max}/\xi'_{\rm F}$, 
can be the same order as $j_{c}$. 

\begin{figure}
\includegraphics[width=6cm,clip]{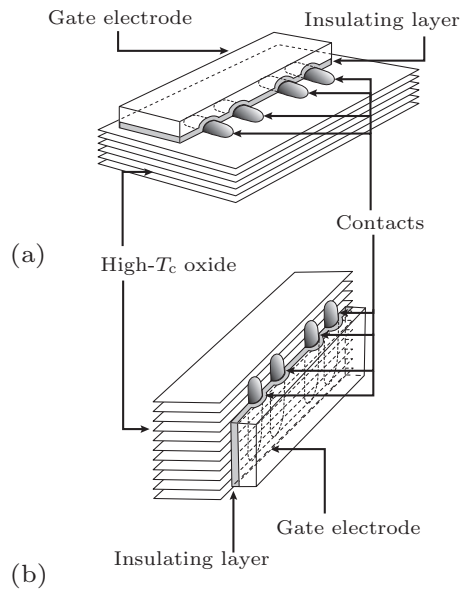}
    \caption{Experimental set-ups using field 
    effect transistor:
    (a) half-gate configuration and 
    (b) side-gate configuration.
    The CuO$_{2}$ planes are shown 
    in an exaggerated scale 
    as compared to the electrodes. 
    }
    \label{FET}
\end{figure}

Finally we point out some experimental set-ups to observe 
the effect predicted in this Letter. 
The simplest way 
may be to utilize the field effect transistor (FET) 
recently developed by Sh\"on {\it et al.} 
\cite{Shoen1,Shoen3}. 
We propose two types of experiments, 
which we call \lq\lq half-gate\rq\rq experiment and 
\lq\lq side-gate\rq\rq experiment. 
The experimental set-ups are depicted in Fig. \ref{FET}. 
In the former case (Fig. \ref{FET}(a)) 
the gate electrode is attached 
parallel to the CuO$_{2}$ plane, 
but it covers only half of it. 
As a result, a boundary between the 
s-RVB ordered region (uncovered by the gate) and 
the hole-rich region (covered by the gate) is 
formed under the edge of the gate. 
(Here we are assuming that the gate 
induces doping to the undoped compound.) 
However, in order to 
observe the effect studied in this Letter,  
the insulating layer 
separating the gate and the sample 
must be smaller than the 
s-RVB coherence length (approximately the 
superconducting coherence length below $T_{c}$
$\sim$ several nm.), 
which may be hardly realized. 

The other configuration is the \lq\lq side-gate\rq\rq 
as depicted in Fig. \ref{FET}(b). 
In this case the gate is applied to the 
CuO$_{2}$ plane from the side 
and by applying a voltage a hole-rich layer is formed 
under the gate. 
This situation is a little different from one considered in 
this Letter where the hole-rich region is infinitely wide. 
However the same mechanism studied by us may also work. 
In this case, the thickness of the superconducting region is 
limited not by the s-RVB proximity length but by the 
thickness of the hole-rich region. 
This configuration can be realized more easily. 

There may be other ways. 
One might introduce the spatial variation of doping chemically, 
for example, by using the epitaxial growth 
\cite{Terashima}. 
If this can be realized, 
it may fit the situation studied in this Letter best. 

In either case, the width of the 
superconducting region is very narrow and 
the fluctuation is essential\cite{Kumagai}. 
Therefore it is not easy to observe 
good superconducting behavior.
However we expect that whether the superconducting correlation 
exists or not is an experimentally detectable fact. 

In summary, we have studied the effect of inhomogeneous 
doping in the high-$T_{c}$ superconductors.  
We have shown, based on the $t$-$J$ model, a possibility of 
superconductivity at the boundary 
between the s-RVB ordered and the hole-rich regions, 
caused by the proximity effect of the s-RVB order. 
We have also suggested experimental set-ups 
to observe this effect. 
Since our proposal is based on the spin-charge 
separation picture, 
these experiments may also shed a renewed light 
on the mechanism of high-$T_{c}$ superconductivity. 
On the other hand, 
from a theoretical point of view, 
the extension of the present study to more 
general inhomogeneous cases is also 
important in understanding recently 
found inhomogeneous phenomena 
in cuprates \cite{Iguchi,Xu,Lang}. 

The authors are grateful to K. Kuboki,  
H. Kohno and M. Mori for many useful discussions. 
This research is supported by a Grant-In-Aid for 
Scientific Research (1246207) from the Ministry of 
Education, Science, Sports and Culture, Japan.


\begin{thebibliography}{99}
\bibitem{average}
This is not exact since now we consider the inhomogeneous doping. 
However rather weak doping dependence of $\xi$ and $\eta$ 
within the mean-field theory 
\cite{Suzumura-Hasegawa-Fukuyama,Kotlier-Liu} 
may justify this approximation.
\end{thebibliography}
\end{document}